\def\gtorder{\mathrel{\raise.3ex\hbox{$>$}\mkern-14mu\lower0.6ex\hbox{$\sim$}}}
\def\ltorder{\mathrel{\raise.3ex\hbox{$<$}\mkern-14mu\lower0.6ex\hbox{$\sim$}}}
\def\eg{{\it e.g.~\hspace*{-1.0mm}}}
\def\ie{{\it i.e.~\hspace*{-1.0mm}}}
\def\etal{{\it et al.~\hspace*{-1.0mm}}}
\def\deg{^\circ}

\def\wv{$V_{606}$~}
\def\wz{$z_{850}$~}
\def\wvv{$V_{606}$}
\def\wzz{$z_{850}$}
\def\viz{{\it viz.~\hspace*{-1.0mm}}}

\documentclass[12pt,preprint]{aastex}
\slugcomment{}

\lefthead{Rix \etal } \righthead{Galaxy Evolution from
Morphologies and SEDs}

\begin{document}
%%%%%%%%%

\title{GEMS Survey Data and Catalog}

\author{
John A. R. Caldwell\altaffilmark{1,2}, Daniel
H. McIntosh\altaffilmark{3}, Hans-Walter Rix\altaffilmark{4}, Marco
Barden\altaffilmark{4}, Steven V.W. Beckwith \altaffilmark{4,5}, Eric
F. Bell\altaffilmark{4}, Andrea Borch\altaffilmark{4}, Catherine
Heymans\altaffilmark{4,6}, Boris H\"au\ss ler\altaffilmark{4}, Knud
Jahnke\altaffilmark{4,7}, Shardha Jogee\altaffilmark{1,8}, Klaus
Meisenheimer\altaffilmark{4}, Chien Y. Peng\altaffilmark{9,1}, Sebastian
F. S\'{a}nchez\altaffilmark{7,10}, Rachel S. Somerville\altaffilmark{1,4}, Lutz
Wisotzki\altaffilmark{7}, Christian Wolf\altaffilmark{11} }
\affil{1 Space Telescope Science Institute, Baltimore MD
21218 USA}
\affil{2 University of Texas, McDonald Obs., Fort Davis TX, 79734 USA}
\affil{3 University of Massachusetts, Dept.~of Astronomy, Amherst MA, 01003 USA}
\affil{4 Max-Planck-Institut f\"ur Astronomie, D-69117 Heidelberg,
Germany}
\affil{5 Johns Hopkins Univ., Dept.~of Physics \& Astronomy, Baltimore MD, 21218 USA}
\affil{6 University of British Columbia, Vancouver, B.C., V6T 1Z1, Canada}
\affil{7 Astrophysikalisches Institut Potsdam, D-14482 Potsdam,
Germany}
\affil{8 University of Texas, Dept.~of Astronomy, Austin TX, 78712 USA}
\affil{9 University of Arizona, Dept.~of Astronomy, Tucson AZ, 85721 USA}
\affil{10 Centro Astron\'omico Hispano-Alem\'an Calar Alto, E-04004 Almer\'ia, Spain}
\affil{11 University of Oxford, Dept.~of Physics, Oxford OX1 3RH, UK {\rm ( \hspace{-1.5mm}}
\footnote{caldwell@astro.as.utexas.edu, dmac@hamerkop.astro.umass.edu,
rix@mpia.de, barden@mpia.de, svwb@stsci.edu, bell@mpia.de, borch@mpia.de,
heymans@physics.ubc.ca, boris@mpia.de, kjahnke@aip.de, sj@astro.as.utexas.edu,
meise@mpia.de, cyp@stsci.edu, sanchez@caha.es, rachel@mpia.de,
lwisotzki@aip.de, cwolf@astro.ox.ac.uk}{\rm \hspace{1.5mm} )}}

\begin{abstract}

We describe the data reduction and object cataloging for the GEMS
survey, a large-area (800 arcmin$^2$) two-band (F606W and F850LP)
imaging survey with the Advanced Camera for Surveys on {\it HST\/}, centered
on the Chandra Deep Field South.

\end{abstract}
 

% In the first two sections, you should notice the use of the LaTeX \cite
% command to identify citations.  The citations are tied to the
% reference list via symbolic KEYs.  We have chosen the first three
% characters of the first author's name plus the last two numeral of the
% year of publication.  The corresponding reference has a \bibitem
% command in the reference list below.
%
% Please see the AASTeX manual for a more complete discussion on how to make
% \cite-\bibitem work for you.

\section{Introduction}
\label{sec:intro}

GEMS ({\bf G}alaxy {\bf E}volution from {\bf M}orphologies and {\bf S}EDs) is a large cycle-11
{\it Hubble Space Telescope} ({\it HST\/}) program
aimed at mapping the evolution of the galaxy population through the 
combination of a large Advanced Camera for Surveys
(ACS) imaging mosaic with ground-based information from COMBO-17 \citep{wolf03}.  \cite{rix04}
gave an overview, and many new results have already been obtained on 
red-sequence galaxies \citep{bell04,mcintosh05,bell05b},
active galaxies \citep{jahnke04,sanchez04}, bar and disk size evolution
\citep{jogee04,barden05}, cosmological weak lensing
\citep{heymans05}, and the cosmic evolution of ultraviolet 
luminosity density and
star formation rate \citep{wolf05,bell05}.
In this paper, we describe in more detail the GEMS data reduction
and master catalog.  The observations and data reduction steps, including
the limiting magnitude
achieved, are discussed in \S 2,3.  The detection and cataloging of objects,
and the correlation of the GEMS source catalog with the COMBO-17 survey catalog 
are described in \S \ref{sec:objects}. In \S 5, we summarize how and where the
GEMS data products can be accessed.

\section{Observations}
\label{sec:obs}

The GEMS collaboration was granted 125 orbits {\it HST\/}
time during cycle 11 (GO-9500, PI: Rix) to image a large
area centered on the Extended Chandra Deep Field South (E-CDFS;
$\alpha_{2000},\delta_{2000}$=03$^h$32$^m$25$^s$,$-27\arcdeg 48\arcmin 50\arcsec$)
using the ACS Wide-field Camera (WFC).
We constructed a tiling scheme (cf.~Figure 1) to obtain $\sim90\%$ coverage of the
$30\arcmin \times 30\arcmin$ E-CDFS region, which had already been surveyed
for rest-frame optical magnitudes and photometric redshifts
by COMBO-17 [``Classifying Objects by Medium-Band Observations in 17
Filters" -- \citep{wolf03}].

The WFC \citep{ford03} consists of a pair of $2048 \times 4096$
pixel ccd detectors separated by a 50 pixel gap. The pixels are $15\micron$,
and the plate scale is $0\farcs05$ pxl$^{-1}$ for an overall extent of
$202\arcsec \times 202\arcsec$.
The gain was set at 1 count per electron, and the a/d
saturation level was 65536 counts.  By virtue of the off-axis, few-element optical
design, affording large area and high throughput, the detector
pixel grid projects to a rhomboidally-distorted grid pattern on the sky, in
which the pixel size varies by a maximum of $19\%$ in a fixed pattern.  During data reduction, this
geometric distortion is removed from the detector image to recover the true sky image.

The GEMS observations consisted of imaging in the F606W and F850LP
passbands, hereafter referred to as \wv and \wzz.  Each {\it HST\/}
visit consisted of three separate 12-13~min exposures each
for \wv and for \wz, dithered by $\sim3\arcsec$ or $\sim60$ pxl in a
three-fold linear spacing which bridges the inter-chip gap
of 50 pxl, and affords some  sub-pixel sampling
of the sky.  In most visits the first orbit observed \wv
and the second one \wzz.   The total exposure times were usually $2160s$ for \wv and $2286s$ for \wzz,
respectively, with the increase reflecting the rapid re-acquisition possible
in the second of two related orbits.  A few observations obtained at a different
spacecraft orientation (tiles 4, 6, 58) yielded integration times of $2286s$ for \wv and
$2160s$ for \wz instead.

Contemporarily with the GEMS observing, the GOODS Project
\citep{giav04} was observing their earliest of five data epochs,
which also used the \wv and \wz passbands.
Figure 1 illustrates the GOODS and COMBO-17 survey areas in relation to
GEMS.  The tiling pattern of the GEMS mosaic was designed to
(a) encompass the 15-tile first epoch GOODS data, (b) create a large contiguous imaging
field, and (c) avoid four extremely bright stars that would risk
charge bleeding and widely scattered light on the detector.  While GEMS-plus-GOODS
corresponds to effectively $9\times9$ tiles, the central and bright star
gaps result in a net of 63 GEMS tile locations, numbered as shown in Figure 1.

The calendar of GEMS data acquisition was as follows.  First observed were tiles
number 6 and 58 in September 2002, with ``orientat" (\viz\/ the
y-axis direction on the resultant images) pointing nearly west ($-92\deg$).  The
bulk of the observations, all but four, were obtained in November 2002
with ``orientat" nearly north ($-2\deg$).  Lastly, tiles number 2 and 4 were
observed in February 2003 with ``orientat" nearly east ($88\deg$).  The
divergent orientations were dictated by guide star availability.

The tiles were chosen with an average overlap of $\sim5\arcsec$ or $\sim100$ pxl,
which resulted in a total data area (including GOODS) amounting
to $28\farcm2 \times 28\farcm2$ (796 sqr-arcmin, or $84.3\%$ of the
$31\farcm5 \times 30\farcm0$ COMBO-17 coverage -- see \S \ref{sec:c17corrl}).  
The guiding stability is quantified by the r.m.s.~guide star pointing
corrections along the two orthogonal symmetry axes of the spacecraft, referred
to as the V2 and V3 directions.
These ``guiding jitter" parameters were in the mean 3.6 and 4.7 mas,
for V2rms and V3rms respectively, with a std.dev.~over all the exposures of
1.1 and 1.0 mas.  This is excellent in relation to the pixel size of 50 mas.  
Note that one tile (chosen to be 44) could have only single-band data, due to the odd
number of orbits.  The execution of the observations
encountered no difficulties and achieved uniformly excellent images.

\section{Data Reduction}
The data were processed in two reduction versions: {\it markI\/}, to provide scientific-quality
data as quickly as possible, and {\it markII\/}, to capitalize on any potential improvements enabled later on.

\subsection{MarkI Reduction}

The calibration of instrumental effects was done by the CALACS ``on the fly"
pipeline \citep{pavlovsky03} as part of the data delivery from the MAST
(``Multimission Archive at Space Telescope") online service.
The CALACS processing subtracts the overscan bias levels,
the superbias image (produced from seven days' intake of bias frames), and the time-scaled superdark
image (produced from one day's intake of dark frames), and then divides by the flatfield (calibrated
by cluster aperture photometry done in orbit).  The ``super" frames achieve
higher statistical accuracy by combining many measurements, yet secular changes
in the detectors, mostly from radiation damage, and practicality dictate a maximum useful time base.
Finally, the gain is corrected to precisely unity, and the FITS header photometry
keyword values are inserted.  In the markI reduction, the data
were requested during August 2002 - February 2003, which invoked
the version 4.1 of CALACS; the August - October 2003 markII reduction
invoked version 4.3.

To remove the geometric distortion, all the exposures for a given tile
and filter were drizzled \citep{fruchter02} onto a celestial pixel
grid centered on the middle step of the \wz three-dither pattern for each
pointing, using a
version of the multidrizzle software \citep{koekemoer03}.  In its
basic function, the multidrizzle task flags bad detector pixels, subtracts the
sky level, drizzles the flux from the detector grid onto the celestial
grid, and then intercompares the result from the separate frames so as
to flag cosmic rays and other transients and find the relative
astrometric corrections to the nominal pointings.  Unweighting all
pixels flagged as invalid, and adding the small astrometric corrections
to the nominal pointings, multidrizzle finally
combines the input frames onto an output image of the weighted average
(over only valid input pixels) counts per second
at each output pixel, which we call the science frame (or tile).
A corresponding weight image is also
produced, namely the effective exposure time contributing at each
output pixel. The adopted weighting scheme disregards the weighting effect from
the amplitude structure of the flatfields and from the inter-frame variations in sky level,
which were judged to be a second order refinement.  Both the registration
and the flagging of invalid pixels can be iterated, working backward from
(\ie\/ re-distorting, also referred to as blotting back) the existing best picture version.
The output image scale was chosen to be $0\farcs30$ pxl$^{-1}$.

Cosmetic blemishes such as
reflection ghosts, diffraction or scattering streaks, or cross-talk
dips were left in the images.  These defects had a sufficiently
limited extent as to have negligible impact on the scientific
usefulness of the images.
The successful removal
of cosmic rays by multidrizzle, without
falsely flagging the centers of real objects, was checked visually.
Its success was facilitated by the property that the data to
be combined were always obtained within a single visit, resulting
in very consistent exposure guiding.  Thus, the estimated pointing
coordinates as stored in the image headers were at least internally
quite consistent between different dithers, although a modest
external correction was needed (cf.~\S 3.3).
A few faint asteroid trails are evident in the
science output and weight images for the tiles \wv 29 and \wz 15 and 48.
They stand out clearly in the weight image if the multidrizzle
software recognized them as transient flux (in effect a long cosmic ray),
but they stand out instead in the science image when they were so weak
as to remain unrecognzied by multidrizzle.  Pixels that were saturated in
the input frames were masked and thus contribute no weight to the final
combined result.

The first epoch GOODS data on the Chandra Deep Field South (CDFS),
consisting of a rectangle of 15 tile positions
(cf.~Fig.~1) at the center of the GEMS array, was analyzed
identically with the markI reduction of GEMS.  The main differences were
that the GOODS \wv data comprised two dither positions at each tile for a
total
exposure time of $1040s$, and the GOODS \wz data comprised four dither
positions
at each tile, for a total exposure time of $2120s$.  Other than distinctions
due to scaling from the somewhat different GOODS exposure times, the first
epoch GOODS data were treated completely interchangeably with the GEMS data, and
will be included as an integral part of them for the rest of the paper.

\subsection{MarkII Reduction}

During the time interval between the markI and markII reductions, because of upgrades to
site-installed software packages, the working version of multidrizzle had evolved.
In addition, we made several changes in our working version of the
software and in the configuration
parameters which control the reduction, based on features in markI that
we thought could be improved upon, 
although the changes turned out to be mostly of a minor or cosmetic nature, and do not imply
any lesser scientific usefulness of the markI quantitative results.
For example, satellite trails, being transients, had been flagged by
multidrizzle in markI but their trail-edge wisps survived this
cleaning procedure.  In markII the program satmask (Richard Hook,
priv.~comm.) was used to pre-excise the entire trail-affected locus
before processing.

The effect of large bright objects upon the sky
subtraction and occasionally also upon the calculated overscan
level, was occasionally noted in markI by some small artificial steps in the calculated
sky-subtracted zero level across the ccd amplifier quadrant boundaries.
By experiment we adopted a better tuning of the sky
level, 0.6(mode) + 0.4(median) over an entire chip (combining the quadrants
having proved more robust).  That still left three of the tiles
with a serious zero level step which was rectified by re-requesting
them under CALACS 4.4.  This last CALACS version had been changed
to remove the earlier vulnerability to extremely bright objects
fortuitously near the ccd chip boundary, which were corrupting the nearby overscan region counts.
The few markII frames needing that one feature from version 4.4 are
otherwise homogeneous with the bulk requested under version 4.3,
since the innovations in 4.4 do not change in any way the resulting
calibrated WFC science array data.

One further innovation in the markII calibration was that CALACS now removed
the differential scale effect due to the changing velocity aberration
between exposures.  Stacking exposures with the differential scale effect could
cause an effective radial blurring (in markI relative to markII), but the
0.04 pxl maximum size of the effect for our data is negligible.

Finally, the drizzling kernel in markII was the lanczos (damped sync function)
\citep{sparks02}, which suppresses one feature of the
correlated noise which results from drizzling, namely the moir\'e
pattern in the noise amplitude, that can be seen for example in
the markI background level; markI used the square drizzling kernel
(pixfrac $=$ 0.8) which was available at the time.
Therefore the noise
pattern in markII is smoother, but the noise remains correlated
due effectively to an irreducible amount of angular averaging that
is inherent in drizzling.
Because of the sensitivity of lanczos algorithm to strong flux gradients, some purely
cosmetic artifacts appeared only in markII at the centers of extremely saturated
stars.  The lanczos kernel furthermore does not behave well when operating
on adjacent pixels with missing data.  We are planning to re-evaluate
the choice of kernel and release the markII reduction publicly only
with the best choice.

\subsection{Astrometric Registration}

The astrometry of each image tile was tied to the overall catalog from
the ground-based COMBO-17 {\it R}-band image \citep{wolf01} using
the wcsfix program (Richard Hook, priv.~comm.).  This relies on
least-squares optimization of the position, orientation, scale, and
axis skew of each tile based on the catalog of objects found
by Sextractor \citep{bertin96}; it is basically identical to the registration
checking of the different dithers within the multidrizzle program
described above.
The \wv frames had a
median of 164 objects bearing on the astrometric registration, with an
$x$ and $y$ registration std.err.~of the mean of 0.21 and 0.20 pxl (at
$0\farcs30$ pxl$^{-1}$).  The \wz frames had a median of 89 useful objects with
an $x$ and $y$ registration std.err.~of the mean of 0.24 and 0.25 pxl, respectively.
The median absolute value size of the shifts needed to correct the instrument pointing
reported in the observation header,
was 55 pxl or $1\farcs6$ in x, 18 pxl or $0\farcs54$ in y, and $0^\circ.010$
of rotation.

Both filters of each GEMS tile are thus tied to the COMBO-17 frame
independently, with an relative uncertainty of typically 0.2 pxl.
The absolute astrometric uncertainty is that of COMBO-17.   To
improve the source color distribution accuracy for proposed applications, a
second version of the V-band frames was generated by {\it
micro-registering} with the IRAF {\it imshift} command and the SExtractored
object position lists, to eliminate
the very small remaining picture shift with
respect to the \wzz-band picture.  This entailed an $xy$-shift with a
uncertainty on the order of 0.02-0.03 pxl at the 0$\arcsec$.03
pxl$^{-1}$ scale.  The effective precision of the \wv registration was
0.0245 pixels or 0.735 mas in the median.

\subsection{Limiting Magnitude}

Because the noise-correlation in the markII-reduced data was more
spatially uniform, these were used to gauge the limiting magnitude of
GEMS.  The edge zone (not covered by all three dithers) 
was pared off and only the lower 90$\%$ of the data
values were used (thus eliminating all objects including outer
haloes).  Then the scatter per point was determined and adjusted
upwards appropriately for the lower 90$\%$ of a normal distribution.
This results in a conservatively high estimate of the scatter, but one
unaffected by objects.  The $5\sigma$ 3 $\times$ 3 input pixels (5
$\times$ 5 output pixels) limiting magnitude was then calculated for
each tile.  A correction of -0.80 magnitude was added to account for
noise correlation based on Casertano \etal (2000) equation (A13).  The
results in Fig.~2 are consistent with a uniform AB limiting magnitude
(5$\sigma$, point sources) of \wv = 28.53 and \wz = 27.27, except for a few tiles with slightly
more scatter.  Tile 56 is affected by a very prominent scattered light
swath from its northwestern bright star.  The other ``hotter" tiles,
2, 4, and 40 may be showing the effect of a small bias subtraction jump
between quadrants.  The range in the limiting magnitude estimates is
quite small.  Further analyses for galaxy fitting, QSO host extraction, bar visibilty,
etc.~will need to make more detailed limiting magnitude
simulations appropriate for their statistical distribution of object
shapes and sizes.

\section{Source Catalog}
\label{sec:objects}
The science goals of GEMS have relied on an empirical database of 
structural and morphological properties for a large and complete sample
of distant galaxies for which redshifts and rest-frame quantities
exist from COMBO-17. Therefore, the initial fundamental steps were the
cataloging of GEMS source detections from the {\it HST} imaging, and
matching these to their counterparts in the COMBO-17 catalog.

\subsection{Object Detection}
\label{sec:detect}
Object detection and cataloging were carried out automatically on the GEMS 
astrometrically calibrated tiles with the SExtractor V2.2.2 software
\citep{bertin96}.  SExtractor identifies sources and provides their
image position, celestial coordinates, projected geometry, and flux parameters,
down to a completeness cutoff that is as uniform as
possible over the area. Moreover, SExtractor produces a segmentation map that
parses the image pixels into those belonging to the extracted sources and the
background sky, which is necessary for fitting galaxy surface brightness
profiles with codes such as GALFIT \citep{peng02} and GIM2D \citep{simard02}.
We have chosen the $0\farcs03$ scale z-band images to be the cataloging basis from consideration
of the science goals.  Before running SExtractor, the science images were converted
from counts-per-second to counts, 
which are related to AB magnitudes via our adopted zero points
ZP[\wvv] = 26.50482 and ZP[\wzz] = 24.84068.  
The conversion to counts is required for the correct SExtractor calculation of
magnitude errors.  

The use of SExtractor in effect defines the extracted sources; thus,
the configuration file must be
carefully tuned for the data set at hand to mininize both crediting noise as spurious objects,
and rejecting believable objects.  As it is common for objects to appear conjoined on the sky due to projection,
further tuning of the configuration file is vital to avoid both the splitting (or ``over-deblending")
of essentially whole objects into pieces, and the lumping together (blending) of different objects
into spurious pseudo-objects.
These considerations are controlled by four configuration parameters:
(1) DETECT\_THRESH, the detecting threshold above background;
(2) DETECT\_MINAREA, the minimum number of connected pixels above threshold;
(3) DEBLEND\_MINCONT, the minimum flux/peak contrast ratio to deblend
separate sources; and (4) DEBLEND\_NTHRESH, the number of deblending
threshold steps.

The primary requirement of our SExtracting was to recover
all $R_{\rm ap}\leq24$ mag galaxies from COMBO-17 with the same spatial
coverage. Fainter than this $R$-band aperture magnitude limit, COMBO-17
photometric redshifts become increasingly unreliable. Even with our
well-defined source detection requirement, we have found that
the large dynamic range of real objects occurring in the data makes it difficult to find
any single configuration parameter set that gives a really satisfactory deblending outcome.  This is
especially so in the present case where the long-exposure ground-based images have reached
low-surface brightness objects, and we are trying to find a match with space-based images that
go especially deep for objects of relatively much smaller angular extent.  To pick up
the $\sim10\%$ of $R_{\rm ap}\leq24$ mag galaxies in COMBO-17 with the
lowest surface brightness, 
the GEMS SExtractor detection threshold would
need to be made so sensitive as to trigger many spurious object detections
on substructural features within, and noise bumps in the outskirts of, 
bright objects (\viz ``over-deblending").
Fortunately, as illustrated in \citet{rix04}, we found that a two-pass 
strategy for object detection and deblending, using separate ``cold'' and
``hot'' SExtractor configurations, met our catalog requirements.

Given the size of the data set, it was essential that any deblending 
method be fully automatic. With this in mind, extensive tests to determine
a combination of two detection configurations, which would maximize the
number of $R_{\rm ap}\leq24$ mag galaxies while minimizing the amount of
overdeblending, were carried out independently by DHM and MB.  Using
five representative \wz frames containing 837 $R_{\rm ap}\leq24$
galaxies, a conservative cold configuration (DETECT\_THRESH$=2.30$, 
DETECT\_MINAREA$=100$, DEBLEND\_MINCONT$=0.065$, and DEBLEND\_NTHRESH$=64$)
identified the high-surface brightness objects with negligible overdeblending,
and a hot version (DETECT\_THRESH$=1.65$, DETECT\_MINAREA$=45$, 
DEBLEND\_MINCONT$=0.060$, and DEBLEND\_NTHRESH$=32$) then extracted
the remaining faint low-surface brightness sources. 
All objects found in the hot run that overlapped the isophotal
area of pre-existing cold run objects, were automatically discounted. For
both configurations a weight map ($\propto{\rm variance}^{-1}$) and a three-pixel
(FWHM) top-hat filtering kernel were used.
The former suppresses spurious
detections on low-weight pixels, and the latter discriminates against noise peaks, which statistically have
smaller extent than real sources as convolved by the instrumental PSF.
The final optimal configurations successfully detected 98.9\% (828/837)
of the COMBO-17 $R_{\rm ap}\leq24$ galaxies with reliable deblending
for 98.1\%.

In addition to the detection parameters, the SExtractor configuration file
includes two parameters used for a global estimate of the background sky
level over a full image. A detailed explanation is provided in
\citet{bertin96}. Briefly, SExtractor constructs a background map by
splitting the image into a grid
of background meshes of a given width (BACK\_SIZE in pixels), applies
a median filter of a given size (BACK\_FILTERSIZE in pixels) to suppress
possible local overestimates, iteratively clips the background histogram
until it converges to within $\pm3\sigma$ of its median value, and
calculates the following mode:
\begin{equation}
{\rm Mode} = 2.5\times {\rm Median} - 1.5\times {\rm Mean} .
\end{equation}
Tests showed that using BACK\_SIZE$=214$ and BACK\_FILTERSIZE$=5$ on
GEMS images provided a reasonable sky level estimation.  The global
background level and the r.m.s. pixel-to-pixel noise
are given for each \wz frame in Table \ref{tab:DetSumm}.

Our final catalog contains 41,681 uniformly and automatically
identified GEMS \wzz-band sources, from 18,528 objects detected in the cold run,
augmented by
23,153 ``good'' objects found in the hot run (after rejection of the 
unwanted hot extensions
to the pre-identified cold objects, described above). The breakdown of
cold, hot, and good sources per ACS frame is given in Table \ref{tab:DetSumm}.
The choice to analyze on a tile-by-tile basis, rather than mosaic-wise, resulted in 
4991 sources detected in multiple overlapping tiles as a result of their
location near image boundaries.  The most interior-located was selected
for entry into the catalog.

The choice of \wz as the primary detection bandpass for cataloging followed from
the importance of the morphology in the reddest bandpass for the science
objectives.  For some applications, \eg weak lensing \citep{heymans05},
we have prepared a \wv-band catalog similarly as described above but with
the hot configuration parameters adjusted as follows:
(DETECT\_THRESH$=1.4$, DETECT\_MINAREA$=37$).
The detectable source density is considerably higher in the \wv images,
and the deblending more problematical due to the greater amplitude of substructure
towards the blue.
Figures \ref{fig3} and \ref{fig4} show the number counts of objects detected 
by GEMS in \wv and \wzz, respectively.

\subsection{Cross Correlation with COMBO-17}
\label{sec:c17corrl}

The GEMS source catalog was cross-correlated in object coordinate space 
with the COMBO-17 source redshift catalog.
For each GEMS source, a
match was accepted for the nearest $R_{\rm ap}\leq24$ galaxy position
within $0\farcs75$. For each tile, we tabulate in Table \ref{tab:DetSumm}
the total number of $R_{\rm ap}\leq24$ mag
galaxies, and the mean and r.m.s.~angular separation $\theta_{\rm sep}$
between the coordinate matches.
There were 1138 cases of ambiguous matching; \ie, unique
COMBO-17 sources with multiple GEMS detections from overlapping images as
described in \S \ref{sec:detect}. These were visually inspected and the
best image detection selected; for 94\% of these, the best case was
the detection farthest from the image edge.  Therefore, our final GEMS and
COMBO-17 cross-correlated catalog contains 8565 $R_{\rm ap}\leq24$ mag 
galaxies, which yielded
an r.m.s.~positional agreement of $0\farcs108$ between the counterparts.
Figure \ref{fig5} shows the good astrometric correspondence between
independently assigned COMBO-17 and GEMS object centroids, which is
encouraging for the key goal of linking space and ground-based information.
This comprises $85.2\%$ of
the 10,056 COMBO-17 sources in the E-CDFS catalog with $R_{\rm ap}\leq24$ mag
and classified as galaxies. The fraction agrees
with the GEMS-to-COMBO area proportion (see \S \ref{sec:obs}), a
further indication that GEMS has reached the goal
of detecting the COMBO-17 sample lying within the 
GEMS-plus-GOODS 796 arcmin$^2$ footprint.

Lastly, Figure \ref{fig6}
compares the relative number of GEMS detections per \wz magnitude bin
that were matched to $R_{\rm ap}\leq24$ mag COMBO-17 galaxies with all detections that
fall into the SExtractor automatic galaxy-like classification
(i.e., CLASS\_STAR$\le0.1$).  We stress that we do not use SExtractor
for star/galaxy separation.  This comparison simply illustrates
that statistically all bright extended sources from the GEMS images are
matched to COMBO-17 redshifts.  The \wz distribution for the 8565 COMBO-17
matches, and for CLASS\_STAR$>0.1$ sources, are given in Figure \ref{fig3}.

\section{Master Catalog and Data Access}

We produced a master catalog in FITS table format, 
which incorporates the GEMS and COMBO-17
information for the 8565 cross-identified objects. This master
catalog was used for the GEMS science completed to date (see \S 1).
A small number of objects were found to be excludable from
the GEMS science sample due to being \eg\/ stars (72), too near the
survey edge (33), or spoiled by the inter-chip gap (62).
These sources are noted in the
master catalog.

The GEMS master catalog and GEMS total source catalog
will be published electronically.  We cite their contents
here by (column) for reference.
For the total GEMS \wz sources: (1) GEMS sexagesimal coordinate-based name, (2-3)
R.A.~and Dec.~(2000), (4-5) total flux (SExtr.~FLUX\_BEST) and error,
(6-7) total magnitude (SExtr.~MAG\_BEST) and error, (8) ``local" background
level, (9) isophotal area (SExtr.~ISOAREA\_IMAGE),
(10-11) image center $x$ and $y$ coordinates (12) position angle
(SExtr.~THETA\_IMAGE),
(13) Ellipticity, (14) image FWHM, (15) SExtractor FLAGS, (16) stellarity
parameter
(SExtr.~CLASS\_STAR), (17-18) ACS source and mask image names, (19) exposure
time, (20) AB-mag zeropoint, (21) SExtractor Kron aperture radius, (22-24)
Cxx Cyy and Cxy SExtractor object ellipse parameters, (25-27) blanks, (28)
number of overlapping sources.  For the master catalog of sources cross
correlated with COMBO-17:
(1-20) as above, (21) COMBO-17 redshift (MC\_z), (22) angular separation
between
ACS and COMBO-17 coordinates, (23) COMBO-17 R-band total magnitude (Rmag),
(24) number of overlapping sources, (25-26) visual-band image $x$
and $y$ coordinates, (27) SExtractor Kron aperture radius,
(28-30) Cxx Cyy and Cxy SExtractor object ellipse parameters,
(31) COMBO-17 object number (Seq), (32) COMBO-17 redshift uncertainty
(e\_MC\_z), (33) COMBO-17 peak of redshift-estimate distribution (MC\_z\_ml),
(34) COMBO-17 (0.3,0.7) luminosity distance (dl),
(35-37) COMBO-17 (0.3,0.7) Johnson M\_U M\_V and M\_B VEGA-mags (UjMag VjMag
and BjMag),
(38) COMBO-17 (0.3,0.7) SDSS M\_r (rsMag), (39) COMBO-17 R-band mag
uncertainty (e\_Rmag),
(40-42) COMBO-17 M\_U M\_V and M\_B mag uncertainties (e\_UjMag e\_VjMag and
e\_BjMag),
(43) COMBO-17 M\_r mag uncertainty (e\_rsMag), (44) COMBO-17 photometry flag
(phot\_flag),
(45) COMBO-17 aperture Rmag (Ap\_Rmag),
(46) COMBO-17 R-band central surface brightness (mu\_max),
and (47) notes.

The MAST archive at the Space Telescope Science Institute makes the GEMS
data products easily available via anonymous ftp; see the instructions at
archive.stsci.edu/prepds/gems.  The GEMS calibrated data so far
stored at the MAST comprise the  
GEMS markI-reduced science and weight images
for both \wv and \wz bands, and also the microregistered version (cf.~\S 3.3)
of the \wv science and
weight images.  The markI-reduced first epoch GOODS science and weight images for
both \wv and \wz bands, and the correspondingly microregistered \wv science
and weight images are also stored there.  A ``readme" file provides
details on the data files and a data summary table provides an overview of
the details specific to each tile, too voluminous to cite here, such as the
exposure date and time, celestial pointing coordinates and sky orientation,
and the total sky level subtracted from combination of images at each
tile location.  The even more detailed accounting for the processing
steps of each ccd chip is contained in the relevant FITS headers.
The GEMS source catalog and the combined GEMS plus COMBO-17 master catalog,
and any revised complete reduction (e.g. markII) will also be placed in MAST.

\acknowledgments

We thank Alison Vick and Guido de~Marchi for support in
preparing the {\it HST\/} observations.  JAC thanks
Mauro Giavalisco (GOODS advice) , Warren Hack and the entire
STScI Computer Support staff (programming help), Richard Hook
(satmask, wcsfix, wdrizzle), Anton Koekemoer (multidrizzle, wcsfix),
and many other STScI colleagues, and lastly the MAST (Multimission Archive at the Space
Telescope) and CADC (Canadian Astronomy Data Centre) for their data
services.

This research was supported by STScI through HST-GO-9500.01.  Support
for the GEMS project was provided by NASA through grant number GO-9500
from the Space Telescope Science Institute, which is operated by the
Association of Universities for Research in Astronomy, Inc.~for NASA,
under contract NAS5-26555.
EFB and SFS acknowledge financial support provided through the European Community's Human Potential Program
under contract HPRN-CT-2002-00316, SISCO (EFB) and HPRN-CT-2002-00305, Euro3D RTN (SFS).
KJ acknowledges support from the Deutsches Zentrum f\"ur Luft- und Raumfahrt (DLR) under project number 50~OR~0404.
SJ acknowledges support from the National Aeronautics and Space Administration (NASA) under LTSA Grant NAG5-13063 issued through the Office of Space Science.
DHM acknowledges support from the National Aeronautics and Space Administration (NASA) under LTSA Grant NAG5-13102 issued through the Office of Space Science.
HWR acknowledges support from the German-Israeli Science Foundation.
CW was supported by the PPARC rolling grant in Observational Cosmology at University of Oxford.

\clearpage

% fig1
\begin{figure}

\plotone{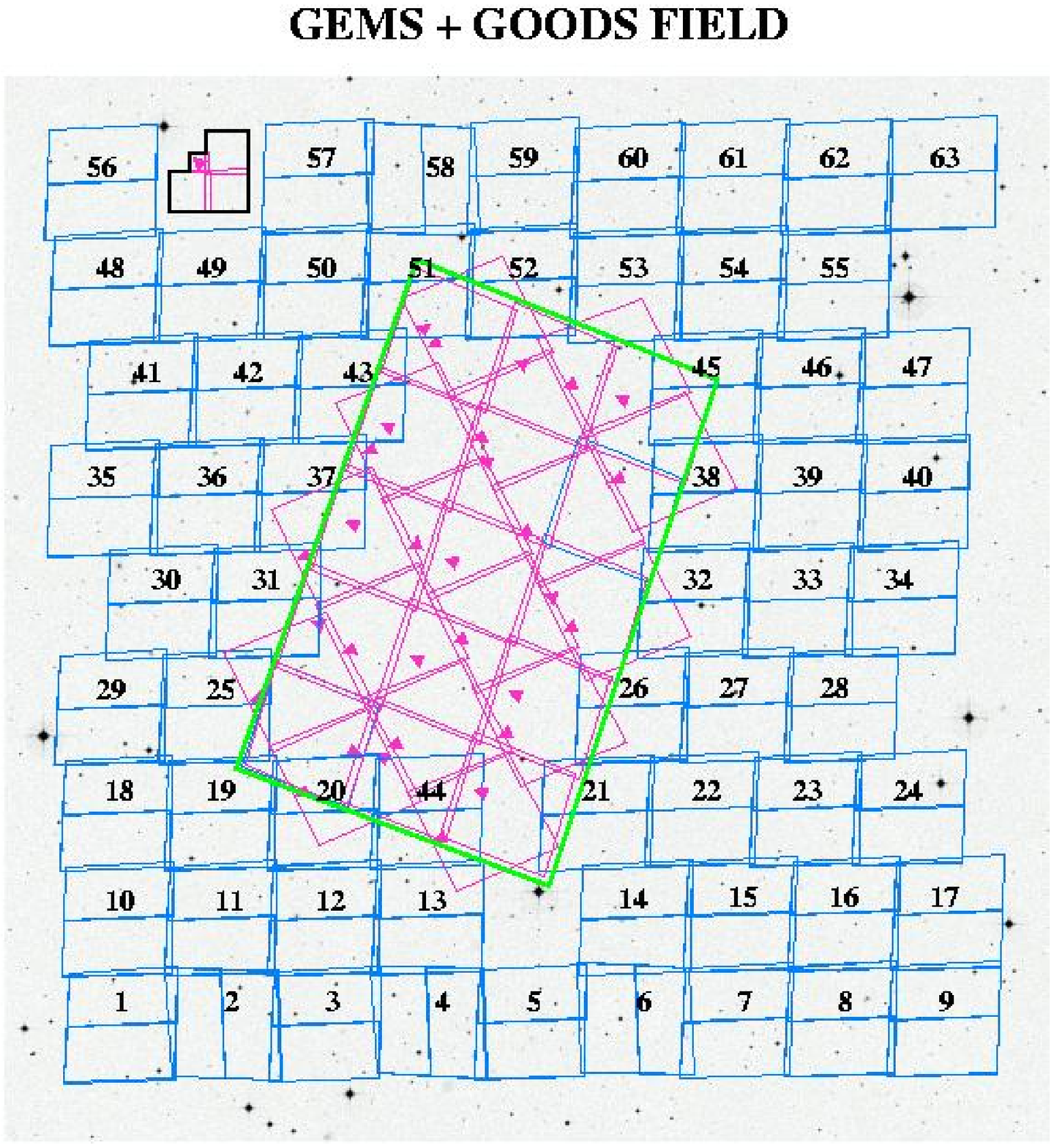} 
\caption{Layout of the GEMS image
mosaic. With 800 arcmin$^2$, GEMS 
nearly covers the extended Chandra Deep Field South from
COMBO-17 (underlying $R$-band image), which measures $\sim
30^\prime \times 30^\prime$; the orientation is North up and East
left. The individual GEMS tiles, labeled by their {\it HST\/} visit
number, are shown as pairs of rectangles (ACS chips). The pink rectangular
mosaic of 15 tiles at center delineates the GOODS first epoch
that has been incorporated in the overall GEMS analysis
(the lapped 16 pink tiles being later epochs and not used).
The tilted
large green rectangle indicates the area of
SIRTF observations for GOODS. The much smaller size of the Hubble
Deep Fields (located elsewhere on the sky) is indicated schematically at top left.  \label{gems-layout}}

\end{figure}

\clearpage

% fig2
\begin{figure}

\plotone{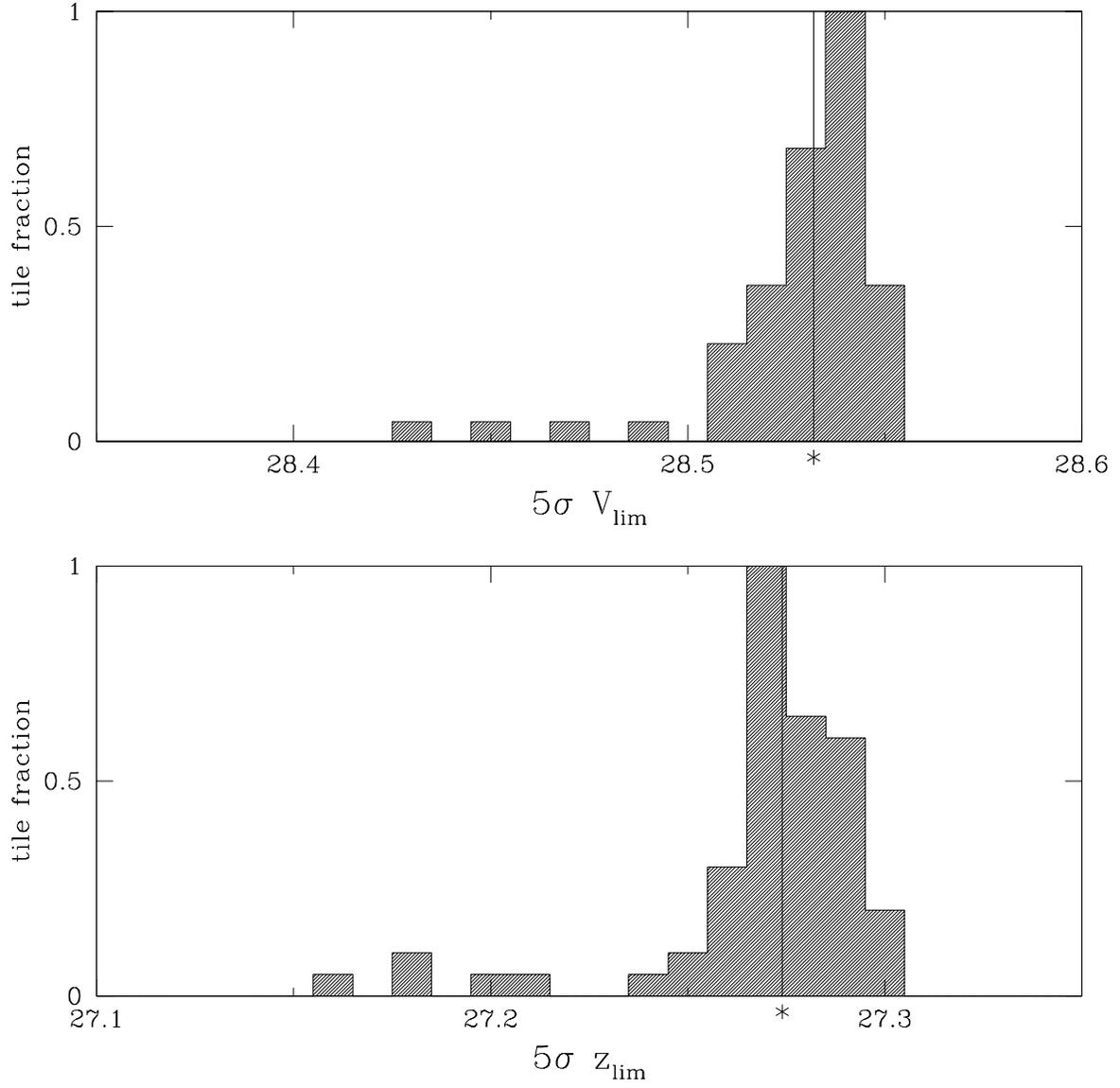} 
\caption{Tile-by-tile estimates of the GEMS
$5\sigma$ limiting magnitude. \label{limmag}}

\end{figure}

% \documentclass[12pt,preprint]{aastex}  %single col., single space
% \usepackage{graphicx,epsfig}
% \begin{document}

\clearpage

% fig3
\begin{figure*}

\center{\includegraphics[scale=1., angle=0]{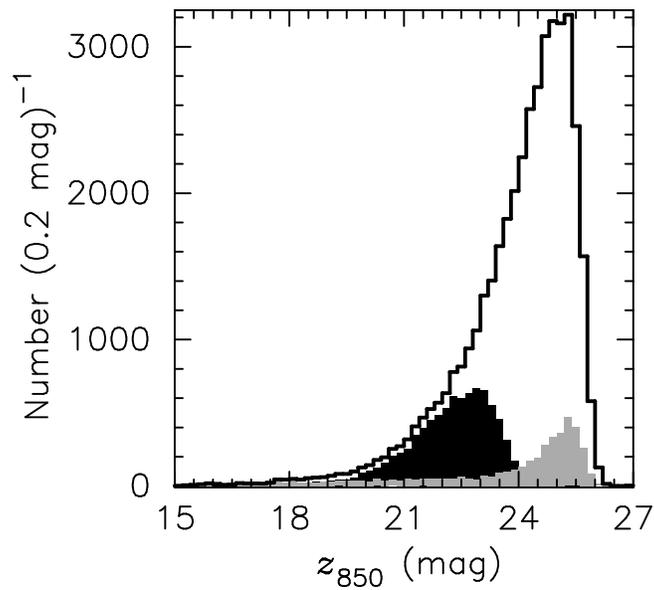}}
\caption[]{\wzz-band source detections in the GEMS {\it HST\/} image mosaic.
For the full set of 78 ACS tiles we plot the 41,681 unique sources 
(black outline) from combining the ``cold'' and ``hot'' configuration SExtractor
catalogs (see text for details).
In addition, we show the subset of
8565 SExtracted sources matched to COMBO-17 galaxies with $R_{\rm ap}\le24$ mag
(black filled distribution), and
the subset of 2251 likely stellar sources with
SExtractor CLASS\_STAR$>0.1$ (gray filled distribution).
\label{fig3}}
\vspace{-0.2cm}
\end{figure*}

\clearpage

% fig4
\begin{figure*}

\center{\includegraphics[scale=1., angle=0]{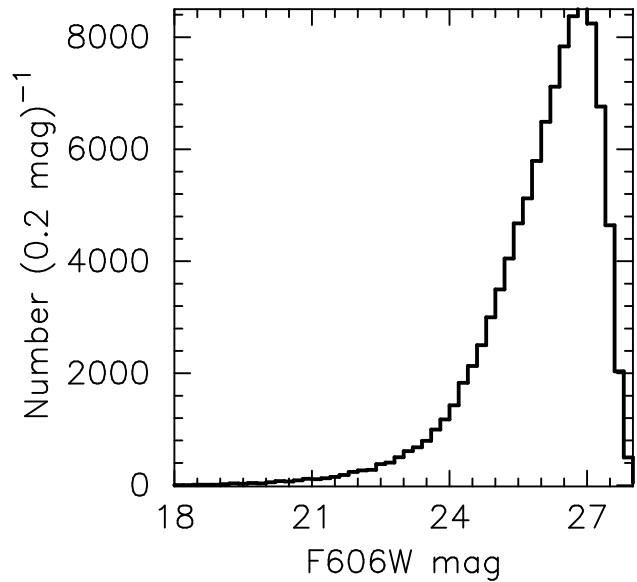}}
\caption[]{Source counts in the GEMS \wv mosaic from 77 ACS images.
We detect 102,138 unique sources using the two-step ``cold'' and ``hot''
SExtractor method, with slightly modified configuration parameters 
\citep[see][]{heymans05}. The increased number of detections compared to the
\wz mosaic (see Figure \ref{fig3}) is the result of the greater
(by 1.2 magnitudes) sensitivity of the \wv imaging.
\label{fig4}}
\vspace{-0.2cm}

\end{figure*}

\clearpage

% fig5
\begin{figure*}

\center{\includegraphics[scale=1., angle=0]{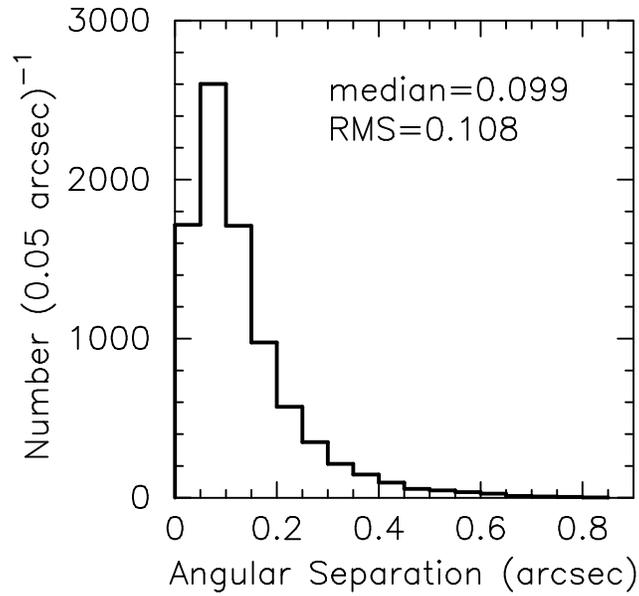}}
\caption[]{Distribution of angular separations between \wzz-band ACS image
and COMBO-17 source positions for the sample of 8565 cross-correlated
galaxies.
We provide the median and root mean square of the distribution.
\label{fig5}}
\vspace{-0.2cm}

\end{figure*}

\clearpage

% fig6
\begin{figure*}

\center{\includegraphics[scale=1., angle=0]{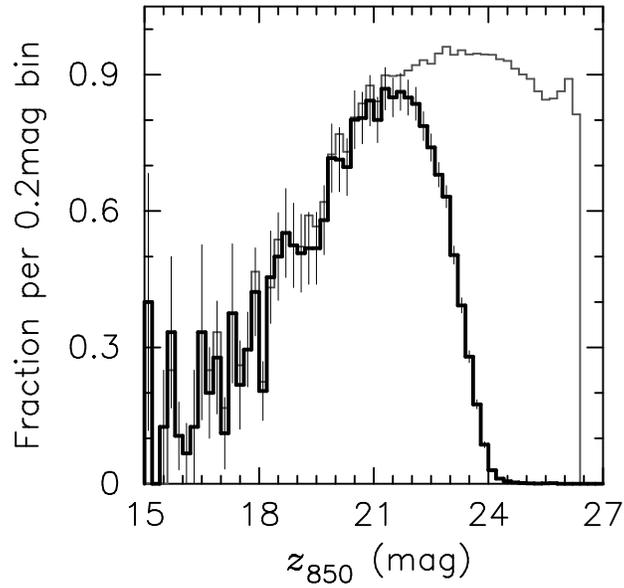}}
\caption[]{Fraction of all \wz SExtractor sources
in GEMS that are known $R_{\rm ap}\le24$ mag galaxies from
COMBO-17 (black with Poisson error bars), and the
fraction that are galaxy-like with SExtractor CLASS\_STAR$\le0.1$
(grey).
\label{fig6}}
\vspace{-0.2cm}

\end{figure*}

%\end{document}

\begin{deluxetable}{lcccccccc}

\tablewidth{0pt}
\tablenum{1}
%tabletypesize{\tiny}
\tabletypesize{\small}
\tablecolumns{9}
\tablecaption{Source Detection Summary}
\tablehead{\colhead{Tile} & \colhead{bkg} & \colhead{$\sigma_{\rm bkg}$} & \colhead{$N_{\rm cold}$} & \colhead{$N_{\rm hot}$} & \colhead{$N_{\rm good}$} & \colhead{$N_{\rm Rap24}$} & \colhead{$\theta_{\rm sep}$} & \colhead{$\sigma_{\theta}$} \\
\colhead{(1)} & \colhead{(2)} & \colhead{(3)} & \colhead{(4)} & \colhead{(5)} &
\colhead{(6)} & \colhead{(7)} & \colhead{(8)} & \colhead{(9)}}
\startdata
GOODS 1  & -0.164 & 4.46 & 173 & 419 & 409 &  84 & 0.10 & 0.10\\
GOODS 2  & -0.155 & 4.48 & 186 & 454 & 446 & 103 & 0.13 & 0.09\\
GOODS 3  & -0.135 & 4.50 & 199 & 430 & 420 & 117 & 0.13 & 0.12\\
GOODS 4  & -0.139 & 4.47 & 226 & 516 & 504 & 117 & 0.13 & 0.11\\
GOODS 5  & -0.161 & 4.50 & 274 & 582 & 563 & 127 & 0.14 & 0.11\\
GOODS 6  & -0.168 & 4.49 & 194 & 451 & 444 & 115 & 0.12 & 0.11\\
GOODS 7  & -0.119 & 4.48 & 233 & 507 & 504 & 140 & 0.12 & 0.11\\
GOODS 8  & -0.101 & 4.46 & 183 & 431 & 421 &  95 & 0.16 & 0.12\\
GOODS 9  & -0.208 & 4.45 & 246 & 552 & 540 & 134 & 0.15 & 0.11\\
GOODS 10 & -0.166 & 4.44 & 242 & 593 & 585 & 151 & 0.15 & 0.12\\
GOODS 11 & -0.113 & 4.48 & 198 & 474 & 461 & 112 & 0.14 & 0.11\\
GOODS 12 & -0.176 & 4.46 & 231 & 534 & 521 & 127 & 0.13 & 0.12\\
GOODS 13 & -0.156 & 4.47 & 263 & 562 & 558 & 138 & 0.15 & 0.10\\
GOODS 14 & -0.206 & 4.45 & 271 & 566 & 560 & 145 & 0.12 & 0.10\\
GOODS 15 & -0.112 & 4.43 & 249 & 537 & 528 & 136 & 0.11 & 0.12\\
GEMS 1  & -0.114 & 3.88 & 235 & 590 & 577 & 102 & 0.14 & 0.12\\
GEMS 2  & -0.169 & 4.28 & 246 & 545 & 536 & 129 & 0.14 & 0.12\\
GEMS 3  & -0.102 & 3.88 & 312 & 679 & 664 & 141 & 0.13 & 0.12\\
GEMS 4  & -0.138 & 4.25 & 250 & 595 & 585 & 139 & 0.13 & 0.12\\
GEMS 5  & -0.075 & 3.90 & 306 & 666 & 639 & 140 & 0.12 & 0.10\\
GEMS 6  & -0.143 & 4.08 & 290 & 601 & 584 & 151 & 0.12 & 0.10\\
GEMS 7  & -0.077 & 3.90 & 285 & 628 & 611 & 115 & 0.12 & 0.12\\
GEMS 8  & -0.172 & 3.92 & 239 & 581 & 558 &  94 & 0.14 & 0.14\\
GEMS 9  & -0.120 & 3.89 & 271 & 614 & 603 & 122 & 0.13 & 0.11\\
GEMS 10 & -0.100 & 3.87 & 239 & 543 & 535 & 106 & 0.12 & 0.11\\
GEMS 11 & -0.113 & 3.83 & 217 & 576 & 570 & 112 & 0.17 & 0.13\\
GEMS 12 & -0.085 & 3.84 & 255 & 674 & 610 & 106 & 0.12 & 0.08\\
GEMS 13 & -0.094 & 3.88 & 236 & 586 & 575 & 130 & 0.14 & 0.13\\
GEMS 14 & -0.079 & 3.92 & 293 & 649 & 631 & 126 & 0.11 & 0.11\\
GEMS 15 & -0.103 & 3.90 & 282 & 613 & 599 & 123 & 0.11 & 0.10\\
GEMS 16 & -0.087 & 3.88 & 297 & 647 & 642 & 124 & 0.13 & 0.12\\
GEMS 17 & -0.152 & 3.87 & 334 & 793 & 768 & 157 & 0.10 & 0.10\\
GEMS 18 & -0.132 & 3.86 & 252 & 639 & 628 & 110 & 0.13 & 0.10\\
GEMS 19 & -0.091 & 3.90 & 248 & 592 & 569 & 104 & 0.16 & 0.12\\
GEMS 20 & -0.108 & 3.86 & 215 & 526 & 515 &  95 & 0.12 & 0.12\\
GEMS 21 & -0.112 & 3.87 & 303 & 700 & 684 & 151 & 0.12 & 0.10\\
GEMS 22 & -0.100 & 3.87 & 307 & 647 & 635 & 140 & 0.12 & 0.11\\
GEMS 23 & -0.113 & 3.82 & 238 & 651 & 636 &  95 & 0.20 & 0.11\\
GEMS 24 & -0.097 & 3.90 & 265 & 648 & 630 & 103 & 0.12 & 0.11\\
GEMS 25 & -0.110 & 3.89 & 299 & 669 & 661 & 131 & 0.14 & 0.11\\
GEMS 26 & -0.102 & 3.89 & 271 & 638 & 623 & 118 & 0.11 & 0.11\\
GEMS 27 & -0.088 & 3.83 & 251 & 638 & 633 & 102 & 0.13 & 0.12\\
GEMS 28 & -0.116 & 3.84 & 268 & 622 & 612 & 113 & 0.12 & 0.12\\
GEMS 29 & -0.213 & 3.92 & 263 & 638 & 612 & 104 & 0.12 & 0.10\\
GEMS 30 & -0.142 & 3.88 & 323 & 736 & 698 & 138 & 0.14 & 0.10\\
GEMS 31 & -0.129 & 3.87 & 282 & 620 & 600 & 126 & 0.14 & 0.10\\
GEMS 32 & -0.145 & 3.88 & 258 & 612 & 606 & 104 & 0.15 & 0.13\\
GEMS 33 & -0.092 & 3.88 & 299 & 703 & 684 & 149 & 0.14 & 0.11\\
GEMS 34 & -0.047 & 3.87 & 277 & 589 & 580 & 111 & 0.12 & 0.11\\
GEMS 35 & -0.108 & 3.91 & 335 & 731 & 714 & 155 & 0.11 & 0.09\\
GEMS 36 & -0.095 & 3.89 & 298 & 666 & 648 & 144 & 0.13 & 0.10\\
GEMS 37 & -0.143 & 3.86 & 237 & 616 & 606 & 116 & 0.13 & 0.12\\
GEMS 38 & -0.120 & 3.84 & 316 & 707 & 693 & 128 & 0.11 & 0.08\\
GEMS 39 & -0.112 & 3.89 & 292 & 734 & 714 & 152 & 0.13 & 0.11\\
GEMS 40 & -0.108 & 3.93 & 292 & 614 & 601 & 140 & 0.17 & 0.12\\
GEMS 41 & -0.178 & 3.99 & 349 & 711 & 698 & 162 & 0.11 & 0.11\\
GEMS 42 & -0.145 & 3.94 & 298 & 644 & 629 & 128 & 0.13 & 0.10\\
GEMS 43 & -0.150 & 3.94 & 316 & 683 & 668 & 145 & 0.13 & 0.13\\
GEMS 44 & -0.050 & 3.89 & 238 & 547 & 535 & 125 & 0.11 & 0.10\\
GEMS 45 & -0.040 & 3.95 & 315 & 712 & 702 & 152 & 0.13 & 0.12\\
GEMS 46 & -0.153 & 3.94 & 295 & 706 & 693 & 126 & 0.15 & 0.11\\
GEMS 47 & -0.168 & 3.92 & 284 & 673 & 657 & 120 & 0.13 & 0.11\\
GEMS 48 & -0.132 & 3.90 & 268 & 653 & 637 & 106 & 0.13 & 0.10\\
GEMS 49 & -0.087 & 3.90 & 243 & 554 & 549 & 101 & 0.14 & 0.11\\
GEMS 50 & -0.116 & 3.96 & 266 & 603 & 580 & 130 & 0.12 & 0.09\\
GEMS 51 & -0.096 & 3.97 & 320 & 705 & 680 & 122 & 0.12 & 0.09\\
GEMS 52 & -0.170 & 3.93 & 298 & 686 & 671 & 112 & 0.15 & 0.11\\
GEMS 53 & -0.139 & 3.92 & 266 & 616 & 601 & 123 & 0.14 & 0.10\\
GEMS 54 & -0.087 & 3.94 & 321 & 723 & 705 & 150 & 0.14 & 0.10\\
GEMS 55 & -0.101 & 3.92 & 280 & 682 & 665 & 123 & 0.13 & 0.11\\
GEMS 56 & -0.332 & 4.06 & 265 & 688 & 612 & 117 & 0.16 & 0.10\\
GEMS 57 & -0.090 & 3.93 & 233 & 582 & 564 &  99 & 0.15 & 0.13\\
GEMS 58 & -0.185 & 4.10 & 287 & 657 & 629 & 134 & 0.15 & 0.13\\
GEMS 59 & -0.080 & 3.90 & 271 & 668 & 649 & 122 & 0.15 & 0.09\\
GEMS 60 & -0.099 & 3.88 & 290 & 657 & 633 & 147 & 0.12 & 0.11\\
GEMS 61 & -0.104 & 3.85 & 295 & 694 & 681 & 119 & 0.14 & 0.11\\
GEMS 62 & -0.098 & 3.85 & 269 & 643 & 632 & 119 & 0.15 & 0.13\\
GEMS 63 & -0.092 & 3.85 & 307 & 693 & 675 & 134 & 0.15 & 0.13\\
Totals  &        &      & 20918 & 48302 & 47078 & 9703 & &   \\
\enddata
\tablecomments{For each \wz GEMS tile listed in (1), we give
the global SExtractor estimate of the background sky level
(2) and the r.m.s.~pixel-to-pixel noise (3).  The source extraction per
tile is summarized by the raw number of ``cold'' (4), ``hot'' (5), and
the combined cold and ``good'' hot (6) detections.
The raw total of $N_{\rm good}=47,078$ includes 4585 duplicate and 406
triplicate source detections (see text for details).
In addition, for each tile we give the number (7) of sources matched to
$R_{\rm ap}\leq24$ mag galaxies from COMBO-17,
and the mean (8) and r.m.s.~(9) angular separation (in arcsec) between
ACS and COMBO-17 coordinates.}
\label{tab:DetSumm}
\end{deluxetable}
\end{document}